\documentclass[12pt, preprint,numberedappendix]{emulateapj}

\newcommand\bibinc{y}		

\usepackage{subeqnarray}
\usepackage{amsmath}
\usepackage{hyperref}
\usepackage{ulem}

\DeclareMathSymbol{\varOmega}{\mathord}{letters}{"0A}
\DeclareMathSymbol{\varSigma}{\mathord}{letters}{"06}
\DeclareMathSymbol{\varPsi}{\mathord}{letters}{"09}

\newcommand{\Eq}[1]{Eqn.~(\ref{#1})}

\newcommand{\Sec}[1]{Section~\ref{#1}}

\newcommand{\Fig}[1]{Fig.~\ref{#1}}

\def \tdrag {\tau_\mathrm{drag}}

\begin{document}

\slugcomment{Draft Modified \today}

\shortauthors{Koll \& Komacek}

\title{Atmospheric Circulations of Hot Jupiters as Planetary Heat Engines}
\author{Daniel D.B. Koll$^{1,3}$ and Thaddeus D. Komacek$^{2,3}$} \affil{$^1$
  Department of Earth, Atmospheric, and Planetary Sciences,
  Massachusetts Institute of Technology, Cambridge, MA, 02139;
  dkoll@mit.edu \\ $^{2}$Lunar and
  Planetary Laboratory and Department of Planetary Sciences, University of Arizona, Tucson, AZ, 85721;
  tkomacek@lpl.arizona.edu \\ $^{3}$ The authors contributed equally
  to this work, and are listed alphabetically.
}

\begin{abstract}
  Because of their intense incident stellar irradiation and likely
  tidally locked spin states, hot Jupiters are expected to have wind
  speeds that approach or exceed the speed of sound. In this work we
  develop a theory to explain the magnitude of these winds. We model
  hot Jupiters as planetary heat engines and show that hot Jupiters
  are always less efficient than
  an ideal Carnot engine. Next,
  we demonstrate that our predicted wind speeds match those from
  three-dimensional numerical simulations over a broad range of
  parameters. Finally, we use our theory to evaluate how well
  different drag mechanisms can match the wind speeds observed with
  Doppler spectroscopy for HD 189733b and HD 209458b.  We find that
  magnetic drag is potentially too weak to match the observations for HD
  189733b, but is compatible with the observations for HD 209458b.
  In contrast, shear instabilities and/or shocks are compatible with both
  observations. Furthermore, the two mechanisms predict
  different wind speed trends for hotter and colder planets than
  currently observed. As a result, we propose that a wider range of
  Doppler observations could reveal multiple drag mechanisms at play
  across different hot Jupiters.
\end{abstract}

\section{Introduction}
\label{sec:intro}

Hot Jupiters provide a unique laboratory for testing our understanding
of planetary atmospheres.  \citet{showman_2002} were the first to
consider the atmospheric circulations of these planets. Using
numerical simulations, \citeauthor{showman_2002} predicted that hot
Jupiters should develop strongly superrotating equatorial jets, with
wind speeds up to several kilometers per second. This prediction was
confirmed by subsequent observations which showed that the thermal
emission peak on many hot Jupiters is shifted eastwards from the
substellar point, consistent with heat being advected downwind by a
superrotating jet \citep[e.g.,][]{Knutson_2007,Crossfield:2010}.

More recent observations have started to directly constrain the wind
speeds of these jets. High-resolution transmission spectra have found Doppler shifts in
molecular absorption lines for HD 209458b \citep{Snellen:2010} as well
as HD 189733b \citep{Wyttenbach:2015,Louden:2015,Brogi:2015}. The
significant ($\sim$ several km s$^{-1}$) blueshifts detected for both
planets imply rapid dayside-to-nightside winds that are broadly
consistent with the wind speeds predicted by a range of numerical simulations
\citep{showman_2002,Showmanetal_2009,Heng:2011a,showman_2013_doppler,Komacek:2017}.

Although it is qualitatively understood why hot Jupiters develop
equatorial jets, there is still no general theory that explains the
jets' magnitude. Hot Jupiters are very likely tidally locked. This
orbital spin state creates a strong day-night forcing which excites
standing waves that flux angular momentum towards the equator and
drive equatorial superrotation \citep{Showman_Polvani_2011}. The
strength of superrotation should therefore depend on the ratio between
horizontal wave propagation and radiative cooling timescales
\citep{Koll:2014,Komacek:2015,Zhang:2016}. This basic expectation is
complicated, however, by results which show that the jet's state
depends on both horizontal standing waves and vertical
eddies \citep{Tsai:2014, Showman:2014}, and it is still unclear how
the two mechanisms jointly determine the jet's magnitude.

In this paper we constrain the wind speeds of hot Jupiters by modeling
their atmospheric circulations as planetary heat engines. The utility
of this approach has previously been demonstrated for hurricanes on
Earth \citep{Emanuel:1986} and rocky exoplanets
\citep{Koll:2016}. Atmospheric circulations can be considered heat
engines because parcels of fluid tend to absorb heat at a high
temperature (e.g., on the dayside of a hot Jupiter) and emit heat at a
low temperature (on the nightside).  The differential heating and
cooling allows parcels to generate work, and thus kinetic energy,
which in steady state has to be balanced by the dissipation of kinetic
energy via friction.

In contrast to hurricanes and the atmospheres of rocky exoplanets,
however, it is still poorly understood how hot Jupiters dissipate
kinetic energy \citep{Goodman:2009}. Potential mechanisms include
magnetic drag in partially ionized atmospheres
\citep{Perna_2010_1,Menou:2012fu,Rauscher_2013,Rogers:2020},
shocks in supersonic flows
\citep{Li:2010,Heng:2012a,perna_2012,Dobbs-Dixon:2013,Fromang:2016},
and turbulence induced by fluid instabilities such as the Kelvin-Helmholtz instability
\citep{Li:2010,Fromang:2016}. 

Our goal is to evaluate these proposed
mechanisms and to test which of them are able to match current
observations. To do so we first describe our numerical simulations
(\Sec{sec:methods}). Next, we develop the heat engine framework and
test it with the numerical simulations (\Sec{sec:theory}). Finally, we
apply our framework to observations (\Sec{sec:data}) and state our
conclusions (\Sec{sec:conc}). Our results
show that current observations favor shear instabilities and/or shocks as the
dominant drag mechanism for HD 189733b, and motivate extending similar
observations across a wider range of planets.

\section{Numerical simulations}
\label{sec:methods}
We compare our theory with the GCM simulations that were previously
described in \cite{Komacek:2017}. In summary, the simulations use the
MITgcm \citep{adcroft:2004} to solve the atmospheric fluid dynamics
equations coupled to double-gray radiative transfer with planetary
parameters relevant for a typical hot Jupiter, HD 209458b.
The double-gray approximation divides the spectrum into an incoming
collimated and a thermal diffuse part. The absorption coefficients
were chosen to match more detailed radiative transfer calculations;
the absorption coefficient for incoming stellar radiation is set to a
uniform value, $\kappa_{SW} = 4 \times 10^{-4}$ m$^{-2}$ kg$^{-1}$,
the thermal absorption coefficient varies approximately with the
square root of pressure, $\kappa_{LW} = 2.28 \times 10^{-6}$ m$^{-2}$
kg$^{-1}$ $\times \ (p/\mathrm{1~Pa})^{0.53}$, where the power-law
exponent comes from fitting the analytic model of
\cite{Parmentier:2014} and \cite{Parmentier:2014a} to radiative
transfer models with realistic opacities. With these values the
photosphere (where the optical thickness equals unity) for stellar
radiation lies at about $0.23$ bar and the photosphere for thermal
radiation lies at $0.28$ bar.
\\ \indent The model's resolution is C32 in the horizontal (roughly
corresponding to a global resolution of $128 \times 64$ in longitude
and latitude) and 40 levels in the vertical which are evenly spaced in
log pressure, with the uppermost layer extending to zero
pressure. Table \ref{table:params} in the Appendix summarizes the
physical and numerical parameters used in our suite of models.

Most GCMs do not explicitly resolve the mechanisms that are thought to
dissipate kinetic energy in hot Jupiter atmospheres, such as Lorentz
drag or shocks (see Section \ref{sec:intro}).  Our GCM
includes two potential sources of drag which can be thought of as
parametrizing these mechanisms.
First, the simulations include a
Rayleigh drag that linearly damps winds over a prescribed timescale
$\tau_{\mathrm{drag}}$. Simulations with
$\tau_{\mathrm{drag}} \leq 10^{5} \mathrm{s}$ use a timescale that is spatially
uniform. Simulations with $\tau_{\mathrm{drag}} > 10^{5} \mathrm{s}$
additionally include a ``basal'' drag term that allows the model to
equilibrate within reasonable integration times. The basal drag
strength increases as a power-law with pressure, from no drag at 10
bar to a timescale of 10 days at 200 bar \citep{Komacek:2015}.
Second,
to enforce numerical stability, the model includes a a fourth-order
Shapiro filter that damps wind and temperature variations at the model
grid scale. The Shapiro filter acts as numerical drag at small spatial
scales and, in simulations without any other sources of drag,
eventually helps to equilibrate the kinetic energy of the flow.
The potential issue with relying on numerical drag is that it relies
on parameters which are generally chosen for modeling
convenience, not because they are physically motivated. This raises
the question of which source of drag is dominant in our simulations.

\begin{figure}
\centering
\includegraphics[width=.45\textwidth]{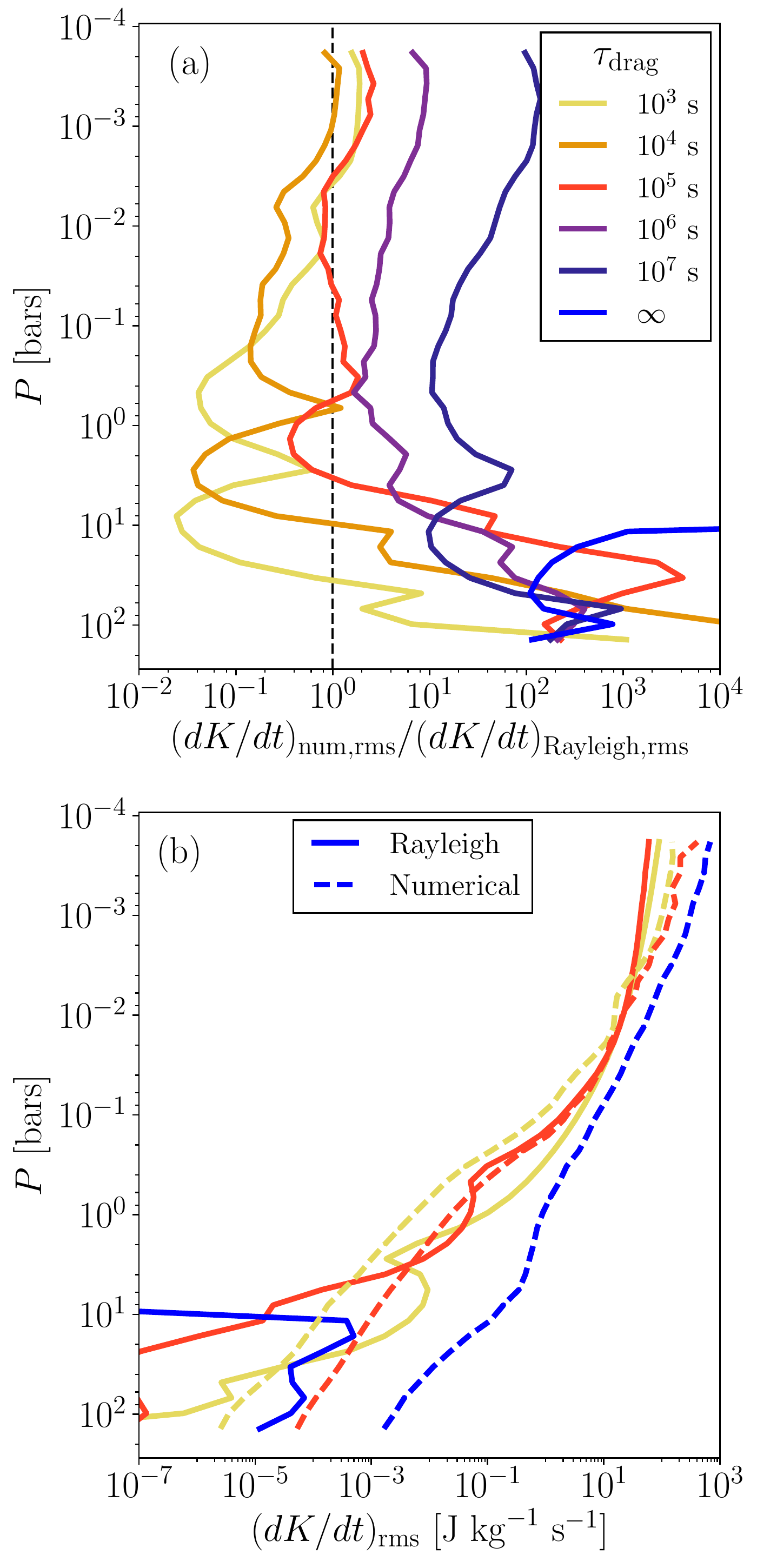}
\caption{ Kinetic energy dissipation in many of our GCM simulations is
  dominated by numerical drag. Panel (a) shows the ratio between the
  global root-mean-square rate of kinetic energy dissipation by
  numerical drag, $(dK/dt)_\mathrm{num,rms}$, versus the global
  root-mean-square rate of kinetic energy dissipation by explicit
  Rayleigh drag, $(dK/dt)_{\mathrm{Rayleigh,rms}}$, as a function of
  pressure for models with an equilibrium temperature of
  $T_\mathrm{eq} = 1500 \ \mathrm{K}$. The colored lines show
  simulations with different Rayleigh drag timescales, with darker
  lines representing longer drag timescales.  The dashed vertical line
  shows the divide between dissipation dominated by numerical drag (to
  the right of the line) and Rayleigh drag (to the left). Except for
  short Rayleigh drag timescales,
  $\tau_\mathrm{drag} \leq 10^{4} \mathrm{s}$, numerical dissipation
  dominates. Note that the case with $\tdrag = \infty$ still includes
  basal drag, so the ratio of numerical to Rayleigh drag dissipation
  is not infinite at depth. Panel (b) shows the absolute contribution
  of Rayleigh drag and numerical effects to the kinetic energy
  dissipation. Only a subset of the simulations are shown for visual
  convenience. The dissipation rate increases with decreasing
  pressure, largely due to the stronger wind speeds at lower
  pressures.}
  \label{fig:dragratios}
\end{figure}

\begin{figure}
\centering
\includegraphics[width=.45\textwidth]{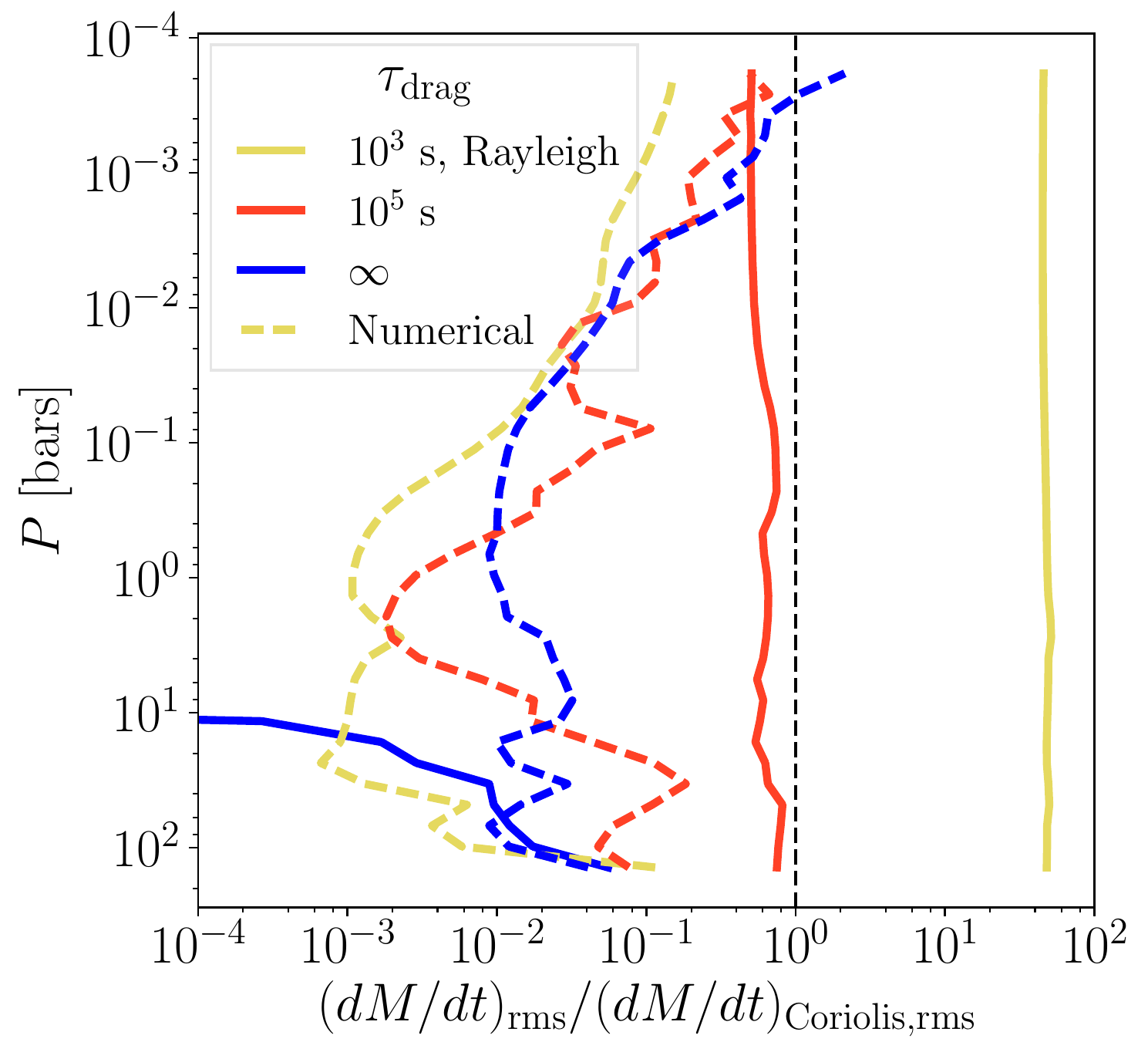}
\caption{Numerical effects are small relative to physical terms
    in the zonal angular momentum budget of our simulations. This plot
    shows the global root-mean-square of the change in zonal momentum due to Rayleigh drag (solid
    lines) and numerics (dashed lines) relative to the change
    in zonal angular momentum due to the Coriolis force
    (i.e. rotation). Plots have the same color scheme as in
    \Fig{fig:dragratios}, for visual convenience we only show a subset
    of all simulations.
    The acceleration from numerics is smaller than either
    the Coriolis force (if $\tdrag \ge 10^5 \ \mathrm{s}$) or Rayleigh drag (if
    $\tdrag \le 10^4 \ \mathrm{s}$). As a result, numerics
    do not significantly affect the angular momentum budget of our
    simulations.}
  \label{fig:angmom}
\end{figure}

We find that numerical drag can play a key role in our GCM
simulations. Although the potential importance of numerical drag has
repeatedly been pointed out in the hot Jupiter literature
(\citealp{Goodman:2009,Li:2010,Thrastarson:2010,Heng:2011a,Liu:2013,Mayne:2014,Polichtchouk:2014,Cho:2015}),
no work has previously quantified its effect relative to explicitly
parametrized drag.
Figure \ref{fig:dragratios} compares the rates at which our GCM is
dissipating kinetic energy via numerical drag from the Shapiro filter
versus the dissipation rate due to Rayleigh drag as a function of
pressure. Figure \ref{fig:dragratios} (a) shows the relative global
root-mean square-dissipation due to numerical drag versus Rayleigh
drag, while Figure \ref{fig:dragratios} (b) shows the absolute global
root-mean-square value of kinetic energy dissipated by both drag
mechanisms. We compute the root-mean-square change in kinetic energy
as
$(\partial K/\partial t)_{rms} = \langle (\partial K/\partial t)^2
\rangle^{1/2}$, where the angle brackets denote an area average.  We
find that all simulations with moderately long Rayleigh drag
timescales, $\tau_{\mathrm{drag}} \geq 10^6 \mathrm{s}$, dissipate
most kinetic energy through numerical drag.

Moreover, even in the simulations with the strongest Rayleigh drag
(yellow curve in \Fig{fig:dragratios}a,b) numerical drag dominates
the dissipation of kinetic energy near
the top and bottom of the model domain.  Although the model includes a
basal drag, we find that it contributes less towards
kinetic energy dissipation than numerical drag near the bottom of the domain. This
is likely due to the Shapiro filter acting as a sponge for waves that
are excited in the upper atmosphere.  However, wind speeds at
pressures greater than 10 bar are small so kinetic energy dissipation
near the domain bottom contributes relatively little to the overall
dissipation (see Fig.~\ref{fig:dragratios}b).

Though numerical drag is a dominant factor in how our GCM
  dissipates kinetic energy, atmospheric circulations additionally
  depend on how the GCM resolves the angular momentum budget. 
  We do not expect \textit{a priori} that numerical effects will
  dominate the global angular momentum budget, because the Shapiro
  filter is designed to not affect large-scale flow \citep{Shapiro:1971}. To
  check this insight, we explicitly compute the change in zonal angular momentum
  by numerics and Rayleigh drag as in \cite{Peixoto:1992}:
\begin{equation}
\label{eq:angmom}
\frac{\partial M}{\partial t} = \frac{\partial u}{\partial t} a \mathrm{cos}(\phi).
\end{equation}
In \Eq{eq:angmom} $M$ is the zonal angular momentum per unit mass, $\partial
M/\partial t$ is the rate of change of angular momentum which we
compute in our simulations from the acceleration $\partial u/\partial
t$ due to the Shapiro filter or Rayleigh drag, $a$ is the planetary radius, and
$\phi$ is latitude.
Rayleigh drag always acts as a sink of angular
momentum whereas the Shapiro filter can accelerate
parts of the flow so we compare both terms via the root-mean-square change in
momentum, $(\partial M/\partial t)_{rms} = \langle (\partial M/\partial t)^2 \rangle^{1/2}$, where the angle brackets as before denote an area average.
\\
\indent We find that numerical effects play a relatively minor role in
the zonal angular momentum budget.
Figure \ref{fig:angmom} shows the change in angular momentum
from numerics and Rayleigh drag relative to the change in angular momentum from the Coriolis force, as a function of pressure. 
We compare both terms against the Coriolis force because it is a
small term in the zonal momentum budget of hot Jupiters due
to their slow rotation and winds that peak at the equator \citep{Showman_Polvani_2011,Showman:2014}.
In relative terms, we find that the numerical change in angular
momentum becomes larger than Rayleigh drag once
$\tdrag > 10^5 \ \mathrm{s}$ (blue curves).  However, in absolute
terms, the momentum change from numerics remains one to two orders of
magnitude smaller than the Coriolis term at most pressure
levels. We conclude that numerical effects
likely do not play a dominant role in the angular momentum budget of
our simulations.

Given that many published simulations of hot Jupiters do not include
Rayleigh drag, our results indicate that many of these simulations rely
on numerical drag to equilibrate kinetic energy. Further work
is needed to ensure that this kind of dissipation in hot Jupiter GCMs
is physically motivated and that its effects are robust with respect
to changes in numerical parameters. At the same time, the angular
  momentum budget in our simulations is not dominated by
  numerics. We therefore expect that GCMs are robust in
  simulating the qualitative features of hot Jupiter circulations
  (e.g., equatorial jets), but that the absolute kinetic energy and
  thus wind speeds in these simulations might be affected by numerical
  details. Our results agree with previous work, which has shown that
  the equilibrated flows in hot Jupiter GCMs largely conserve angular
  momentum, are independent of initial
  conditions, and that the magnitude of winds is only weakly sensitive
  to changes in numerical parameters
  (e.g. \citealp{Heng:2011,Liu:2013,Mayne:2014}). In the
  remainder of this paper we focus on existing GCMs to test our
theoretical framework. To do so we develop a theory in the next
section that can account for both explicit and numerical drag.

\section{Hot Jupiters as heat engines}
\label{sec:theory}

In steady state, the rate $W$ at which a heat engine performs work is
given by
\begin{equation}
W = \eta Q,
\label{eq:carnot}
\end{equation}
where $\eta$ is the engine's thermodynamic efficiency and $Q$ is the
rate at which the engine absorbs heat.

First, the heating rate $Q$ is equal to the average absorbed
stellar flux,
\begin{equation}
\label{eq:dotq}
Q = \sigma T_{eq}^4,
\end{equation}
where $T_{eq}$ is the planetary equilibrium temperature.

Second, we constrain the work output rate $W$.  We assume that work
goes entirely towards generating and dissipating kinetic energy.
If Rayleigh drag dominates, the rate at which kinetic energy is
dissipated equals
\begin{equation}
\label{eq:dotwray}
W_\mathrm{Rayleigh} = \int \frac{dp}{g} \times \left\langle \frac{\mathbf{v}^2}{\tau_{\mathrm{drag}}}\right\rangle \mathrm{,}
\end{equation}
where $\mathbf{v}$ is the velocity vector and the angle brackets
denote an area average.
If numerical drag dominates, kinetic energy is dissipated by the
Shapiro filter which damps the highest wavenumber components of the
flow. Because the highest wavenumber in the GCM is set by the model's
grid spacing $\Delta x$ we scale the Shapiro filter's damping
timescale as $\tau \sim \Delta x/ U$. This means the rate at which
numerical drag dissipates kinetic energy is equal to
\begin{equation}
\label{eq:dotwnum}
W_\mathrm{num} \sim \frac{U^2}{\Delta x/ U} \times \frac{p}{g} =
\frac{U^3}{\Delta x} \times \frac{p}{g}.
\end{equation}

\begin{figure}
\includegraphics[width=0.5\textwidth]{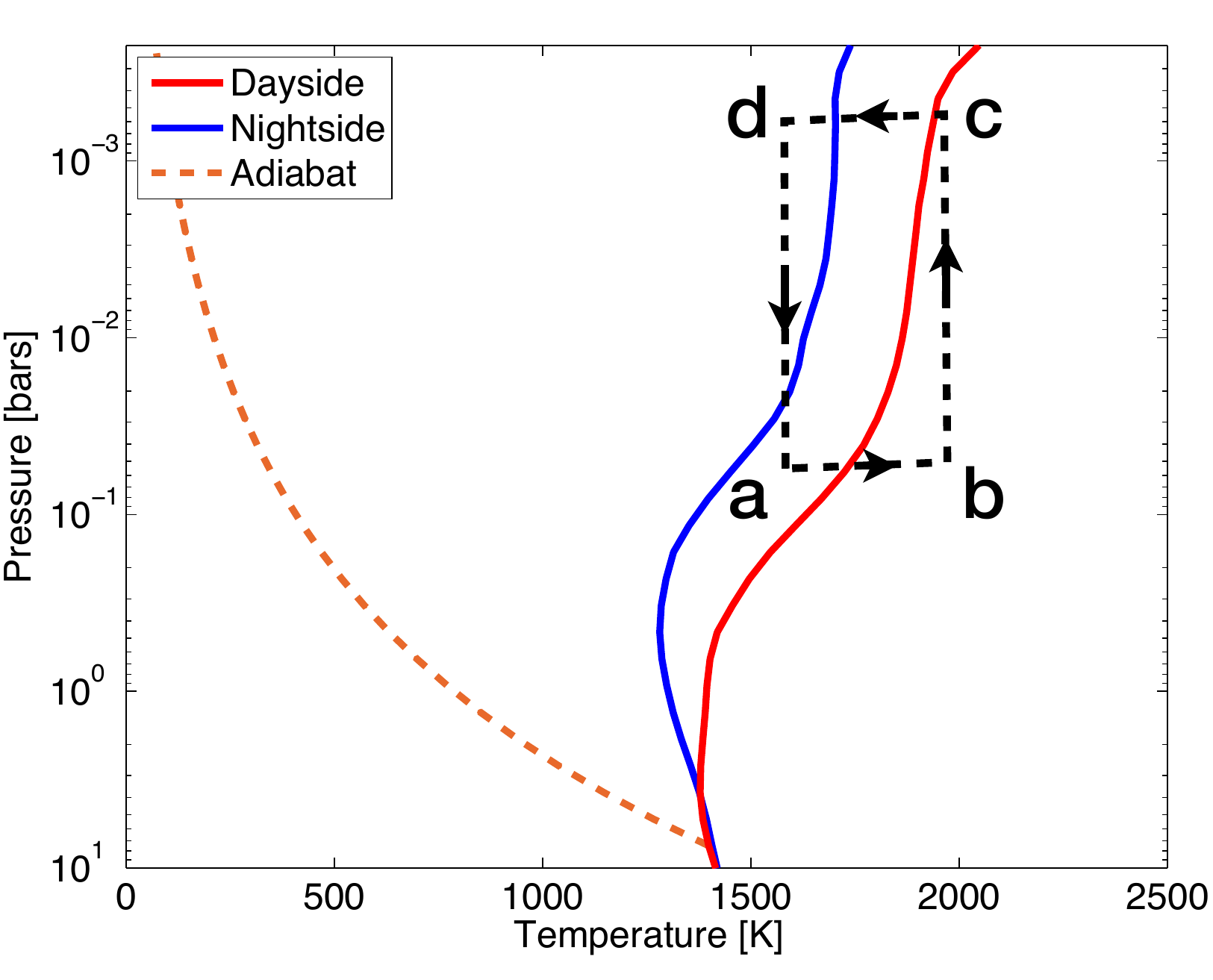}
\caption{
  A diagram of the Ericsson cycle, overlaid on dayside- and nightside-averaged
  temperature profiles of a reference simulation ($T_\mathrm{eq} =
  1500 \ \mathrm{K}, \tau_\mathrm{drag} = 10^6 \ \mathrm{s}$) and an
  adiabatic profile.
  The Ericsson cycle works as follows: a parcel of fluid starts at depth on the
  nightside (a), moves towards the dayside (b), where it rises (c),
  moves back towards the nightside (d), and sinks (a). We assume that
  rising and sinking motions (b-c, d-a) are isothermal and that
  motions between hemispheres (a-b, c-d) are isobaric.
  The isothermal assumption is motivated by the GCM profiles, which
  show that hot Jupiters are much closer to vertically isothermal than
  to adiabatic.}
\label{fig:diagram}
\end{figure}

Third, we constrain the efficiency $\eta$. Previous work on
  hurricanes and the atmospheres of rocky planets constrained this
  quantity by modeling atmospheric circulations as Carnot cycles
  \citep{Emanuel:1986, Koll:2016}. Unfortunately it is difficult to
  argue that hot Jupiters should also resemble Carnot cycles.  In a
  Carnot cycle parcels of fluid expand and contract adiabatically
  between heating and cooling. This model is physically motivated by
  the fact that hurricanes and rocky planets undergo convection, so
  fluid parcels move rapidly and quasi-adiabatically. In contrast, the
  upper atmospheres of hot Jupiters are strongly irradiated by their
  host stars. The irradiation creates a stable stratification and
  suppresses convection, which means the vertical temperature
  structure is approximately in radiative equilibrium and lapse rates
  are small \citep{Iro:2005,Guillot:2010}. As the temperature profiles
  from a reference simulation in Figure \ref{fig:diagram} illustrate,
  temperatures are indeed far from adiabatic, which underlines that
  the Carnot cycle is a poor model for hot Jupiters.

Here we constrain the efficiency $\eta$ by modeling hot Jupiters as
Ericsson cycles \citep{McCulloh:1876}. The Ericsson cycle is shown in
Figure \ref{fig:diagram}: a parcel of fluid starts deep in the
nightside atmosphere (Fig.~\ref{fig:diagram}, point a). It moves at
constant pressure towards the dayside (b), where the stellar heating
causes it to rise (c). The parcel then moves to the nightside (d),
before cooling and sinking back to its starting position (a).
Even though the assumption of isothermal vertical motions is an
idealization, Figure \ref{fig:diagram} shows that the Ericsson cycle
provides a physically motivated model for hot Jupiters.

The efficiency of the Ericsson cycle is given by 
\begin{equation}
\label{eq:etafrac}
\eta = \frac{\oint \delta Q}{\int_a^c \delta Q} = \frac{\oint T ds}{\int_a^c T ds} \mathrm{.}
\end{equation}
Here
$\delta Q$ is a change in a parcel's heat content, and $ds$ is a change in entropy. From the
first law of thermodynamics, 
\begin{equation}
T ds = c_p dT - \frac{dp}{\rho} = c_p dT - R Td\ln p \mathrm{,}
\end{equation}
where we have used the ideal gas law in the second step. We can then
evaluate the numerator $\oint T ds$ as
\begin{eqnarray}
&& \int_a^b c_p dT - \int_b^c R T \nonumber
d\ln p + \int_c^d c_p dT - \int_d^a R T d\ln p, \\
  & = &  c_p (T_\mathrm{day} - T_\mathrm{night}) - R T_\mathrm{day} \ln(p_\mathrm{lo}/p_\mathrm{hi})
        \nonumber \\ &&  +  c_p
        (T_\mathrm{night} - T_\mathrm{day})  - R T_\mathrm{night} \ln(p_\mathrm{hi}/p_\mathrm{lo}), \nonumber \\
  & = & R (T_\mathrm{day} - T_\mathrm{night}) \ln(p_\mathrm{hi}/p_\mathrm{lo}).
\end{eqnarray}
Similarly the denominator $\int_a^c T ds$ in \Eq{eq:etafrac} is
\begin{eqnarray}
&& \int_a^b c_p dT - \int_b^c R T d\ln p, \nonumber \\
  & = & c_p (T_\mathrm{day} - T_\mathrm{night}) + R T_\mathrm{day} \ln(p_\mathrm{hi}/p_\mathrm{lo}).
\label{eq:integraltwo}
\end{eqnarray}
The ratio of these two terms gives the efficiency, which we
write as
\begin{equation}
  \eta = \frac{ \frac{T_\mathrm{day} - T_\mathrm{night}}{T_\mathrm{day}} \times 
  \ln\left[ (p_\mathrm{hi}/p_\mathrm{lo})^{R/c_p} \right]}{ \frac{T_\mathrm{day} -
    T_\mathrm{night}}{T_\mathrm{day}} + \ln\left[ (p_\mathrm{hi}/p_\mathrm{lo})^{R/c_p}\right]}.
\label{eq:eta}
\end{equation}

\begin{figure*}
\centering
\includegraphics[width=0.8\textwidth]{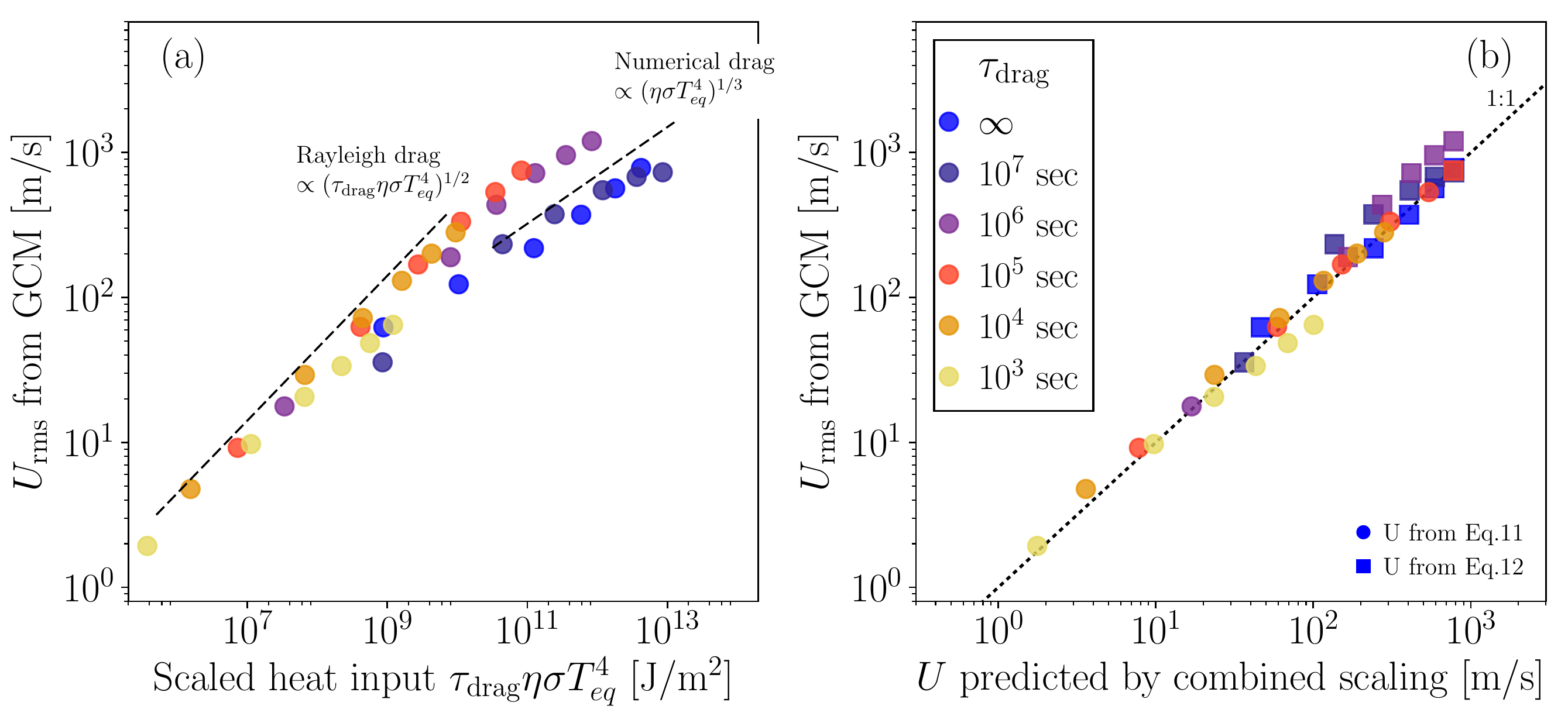}
\caption{ Our heat engine scaling captures the strength of wind speeds
  across a wide range of hot Jupiter GCMs. (a) Hot Jupiter
    simulations fall into two regimes in which bulk wind speeds
    either scale following Rayleigh drag or numerical
    drag (black lines show the two different slopes). (b) Our combined scaling predicts the GCM
    wind speeds in both regimes.  The y-axis corresponds to the
  root-mean-square wind speed averaged over pressures less than
  $1 \ \mathrm{bar}$ in different hot Jupiter simulations, the
    x-axis is the wind speed predicted from Equation
    \ref{eq:Ucombined}. Each dot represents a different GCM
  simulation with varying $T_\mathrm{eq} = 500 - 3000 \ \mathrm{K}$,
  where different colors represent different Rayleigh drag timescales
  used in the simulations. The black line indicates a 1:1 fit between
  theory and simulations. Circles show simulations for which
    we use the Rayleigh drag scaling (Eqn.~\ref{eq:Uray}), squares
    show simulations for which use the numerical drag scaling
    (Eqn.~\ref{eq:Unum}). Note: To display
    simulations without Rayleigh drag (blue dots), for which
    $\tau_{drag}=\infty$, we use $\tau_{drag}=5\times10^6$s in the
    left panel instead.}
\label{fig:winds}
\end{figure*}

Importantly, the efficiency $\eta$ is always
  lower than the efficiency of a Carnot cycle,
$\eta_{\mathrm{Carnot}} = (T_\mathrm{day} - T_\mathrm{night}) /
T_\mathrm{day}$, which is the maximum efficiency a heat engine can
reach.  The lower efficiency arises because heat
is radiated to space as a parcel passes from the dayside to the
nightside (c-d). If, instead, this heat could be stored and used later
to heat up the parcel as it passes back from the nightside to the
dayside (a-b), the Ericsson cycle's efficiency would equal that of a
Carnot cycle\footnote{If the heat lost during (c-d) could be captured
  and used to heat the parcel during (a-b), then \Eq{eq:integraltwo}
  becomes
  $\int_a^c \delta T ds = \int_b^c \delta T ds = R T_\mathrm{day}
  \ln(p_\mathrm{hi}/p_\mathrm{lo})$ and \Eq{eq:eta} becomes
  $\eta = (T_\mathrm{day} - T_\mathrm{night}) / T_\mathrm{day}$.}.
%

As an example we consider the efficiency of WASP-18b, whose
phase curve is consistent with zero heat redistribution from dayside
to nightside \citep{Maxted:2013}. We assume that a parcel of fluid
moves two scale heights in the vertical every time it traverses the
planet horizontally\footnote{A parcel travels a vertical distance
  $d_\mathrm{z} \sim \frac{Wa}{U}$, where $a$ is the planet radius and
  $W$ the vertical wind speed. Using characteristic values from a
  simulation with $T_\mathrm{eq} = 1500 \ \mathrm{K}$ and no drag,
  $W \sim 10 \mathrm{m}\mathrm{s}^{-1}$,
  $U \sim 10^3 \mathrm{m}\mathrm{s}^{-1}$, and
  $a=a_\mathrm{HD 209458b}$, we find $d_\mathrm{vert} \sim 2H$, where
  $H$ is the scale height. In agreement with this estimate, we mapped
  streamfunctions in our simulations and found that the vertical
  extent of both zonal and meridional flows is normally confined to
  $\sim 1-3$ scale heights.} so
$\ln\left[ (p_\mathrm{hi}/p_\mathrm{lo})^{R/c_p} \right] \sim 2
R/c_p$. In this case WASP-18b's Carnot efficiency would be unity,
$\eta_\mathrm{Carnot} = 1$, whereas its actual efficiency is smaller
by a factor of three, $\eta = 0.36$.
Hot Jupiters can therefore be thought of as comparable to, but
  less efficient than, ideal Carnot engines.
Their efficiency can be reduced even further by molecular
diffusion and irreversible phase changes \citep{pauluis2002a}, so
Equation \ref{eq:eta} should be considered an upper limit.

We are now able to test the extent to which hot Jupiters
  resemble heat engines. A key prediction of our theory is that wind
  speeds are sensitive to whether winds are damped by Rayleigh drag or
  numerical drag.  Based on Equations
  \ref{eq:carnot}-\ref{eq:dotwnum}, we expect that winds should scale
  as the square root of the modified heat input for Rayleigh drag,
  $U \propto (\tau_{\mathrm{drag}} \eta \sigma T_{eq}^4)^{1/2}$,
  whereas they should scale as the one-third power of the heat input
  for numerical drag,
  $U \propto (\Delta x \eta \sigma T_{eq}^4)^{1/3}$.
  To compare both scalings in a single plot and because $\Delta x$
  depends on numerical parameters we first use the quantity
  $\tau_{\mathrm{drag}} \eta \sigma T_{eq}^4$.
  \\ \indent Figure \ref{fig:winds}(a) shows that our simulations indeed
  exhibit a dichotomy between Rayleigh and numerical drag. The x-axis
  shows the scaled heat input
  $\tau_{\mathrm{drag}} \eta \sigma T_{eq}^4$ while the y-axis shows
  the root-mean-square wind speed,
  $U_{rms} = (p^{-1} \int \langle u^2 + v^2 \rangle dp)^{1/2}$, where
  $u$ and $v$ are the zonal and meridional wind speeds and where we
  average horizontally and over the meteorologically active region
  above $p=1$ bar (see Fig.~\ref{fig:dragratios}).  To evaluate $\eta$ we use the dayside and
  nightside brightness temperatures that would be seen by an observer
  and assume that a parcel crosses two scale heights,
  $\ln\left[ (p_\mathrm{hi}/p_\mathrm{lo})^{R/c_p} \right] \sim 2
  R/c_p$.
  \\ \indent We find that wind speeds in most strongly damped
  simulations with $\tau_\mathrm{drag} \leq 10^5 \mathrm{s}$ increase
  according to Rayleigh drag (Fig.~\ref{fig:winds}a). In contrast,
  winds in simulations with $\tau_\mathrm{drag} \geq 10^6 \mathrm{s}$
  increase more slowly and approximately follow the one-third slope
  predicted for numerical drag.
  A notable exception to the Rayleigh scaling is given by the 
  hottest simulations with $\tau_\mathrm{drag}=10^3 \mathrm{s}$
  (yellow dots), in
  which winds increase with a one-thirds slope instead.
  This is due to the relative increase of numerical dissipation in
  strongly damped simulations. At $\tdrag = 10^3 \mathrm{s}$
  winds are so weak that Rayleigh drag, which is proportional to
  wind speed, becomes small relative to numerical drag in parts
  of the model domain.
  Similarly, our numerical scaling performs worst for simulations with
  $\tau_\mathrm{drag}=10^7 \mathrm{s}$ (purple dots), in which wind speeds flatten out at
  high $T_{eq}$ even though the heat input keeps increasing.
  Given that our theory performs well in the strongly damped limit,
  deviations from it are likely due to inaccuracies in
  our numerical scaling, which we discuss below.

  We now constrain the wind speeds inside a hot Jupiter atmosphere. If
  the atmospheric circulation is primarily balancing Rayleigh drag
  then wind speeds should scale as
\begin{equation}
U_\mathrm{Rayleigh} = k_0 \left(\tau_\mathrm{drag} \eta
  \sigma T_{eq}^4 \frac{g}{p} \right)^{1/2},
\label{eq:Uray}
\end{equation}
whereas if the circulation is balancing numerical drag then winds should scale as
\begin{equation}
U_\mathrm{num} = k_1 \left(\Delta x \eta \sigma T_{eq}^4 \frac{g}{p} \right)^{1/3}.
\label{eq:Unum}
\end{equation}
Here $k_0$ and $k_1$ are fitting constants of order unity that account
for various approximations, in particular our assumption that
temperature profiles are isothermal. We use $k_0=0.3$ and
  $k_1=1.1$ to match the simulations at $T_{eq} = 3000$ K with
  $\tau_\mathrm{drag} = 10^4 \mathrm{s}$ and $\tau_\mathrm{drag} = \infty$,
  respectively.
We combine Eqns. \ref{eq:Uray} and \ref{eq:Unum} by demanding that a
GCM's work output equals whichever is stronger, Rayleigh or numerical
drag, so
\begin{equation}
U = \min(U_\mathrm{Rayleigh}, U_\mathrm{num}).
\label{eq:Ucombined}
\end{equation}
To evaluate \Eq{eq:Unum} we use the model's grid spacing at
the equator $\Delta x \sim 2 \pi a/128$, where $a$ is the planetary
radius.

We find that our theory matches the GCM simulations
well. Figure \ref{fig:winds}(b) compares our
predicted winds with the simulated
root-mean-square wind speeds $U_{rms}$, defined above.
As in Figure \ref{fig:winds}(a), we find that our
scaling works best in the strongly
damped limit, particularly for the simulations with
  $\tau_\mathrm{drag} = 10^4-10^5 \mathrm{s}$ which our scaling
  matches to better than $33\%$. These are also the simulations in
which numerical drag is not dominant yet, and for which we scale winds using
\Eq{eq:Uray}.

Our scaling additionally matches the weakly damped simulations that
are dominated by numerical drag
($\tau_\mathrm{drag} > 10^5 \mathrm{s}$), even though the fit is less
good than in the strongly damped regime. This is likely due to the
approximations we made in deriving \Eq{eq:Unum}. To test this point we
performed additional simulations in which we varied the model
resolution and timestep. We found that \Eq{eq:Unum} over-predicts the
sensitivity of wind speeds to numerical resolution (see Appendix).
Further work is needed to understand exactly how hot Jupiter
simulations equilibrate through numerical drag.

Nevertheless, given that our scaling captures the basic dependence of
wind speeds on a planet's heat input (Fig.~\ref{fig:winds}a) and
additionally matches the GCM to better than a factor of two even when
the models are dominated by numerical drag (Fig.~\ref{fig:winds}b), we
argue that the main shortcoming in Figure \ref{fig:winds} is due to
our imperfect description of numerical drag, not due to the heat
engine framework.
We therefore sidestep the intricacies of numerical simulations
and in the last section apply the heat engine framework directly to
data.

\section{Evaluating drag mechanisms with observations}
\label{sec:data}

\begin{figure}
\includegraphics[width=.5\textwidth]{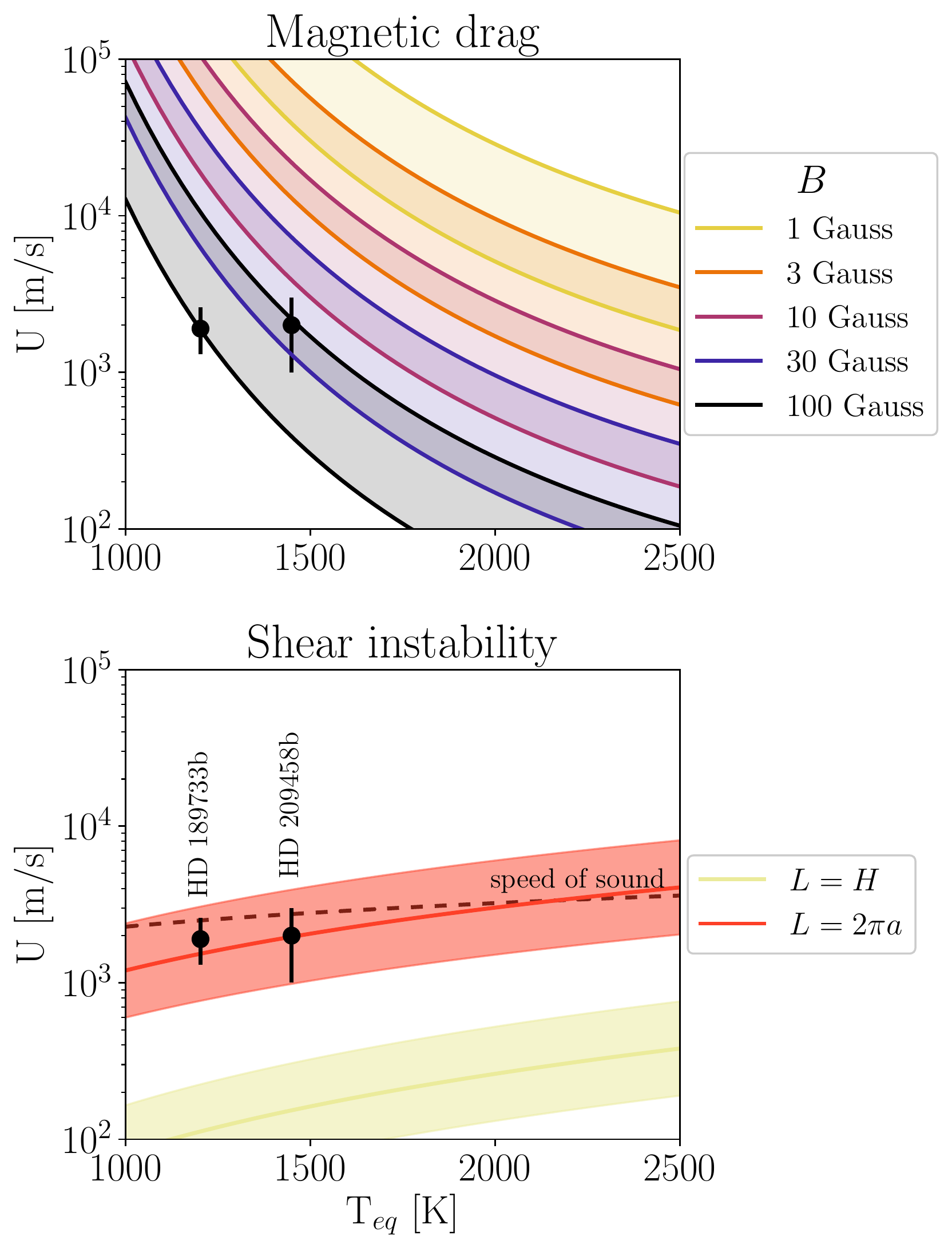}
\caption{Top: solid lines show the predicted wind speeds from
  \Eq{eq:Uray}, assuming dissipation is caused by magnetic
  drag. Colored envelopes indicate that our theoretical
    scalings are subject to uncertainty. The uppermost line for each
    magnetic field strength shows the wind speed predicted for
    dissipation occuring at $1$ bar, the lower line shows the wind
    speed predicted for dissipation occuring at $10^{-3}$ bar, and the
    colored envelope shows intermediate pressures. Dots show wind
  speeds constrained via Doppler spectroscopy for HD 189733b and HD
  209458b \citep{Snellen:2010, Louden:2015}. Bottom: solid lines show
  the predicted wind speeds from \Eq{eq:shearU}, assuming dissipation
  is caused by shear instabilities. Colored envelopes here
    indicate our estimated uncertainty for our heat engine scaling
    (see text). Winds faster than the speed of sound (dashed black
  line\footnote{We assume solar composition and that atmospheric
    temperature is equal to the equilibrium temperature.}) can also
  develop shocks.  %
  Magnetic drag can match both observations, but doing so requires
    a large dipole field ($\gtrsim 100 \mathrm{G}$) for HD 189733b.
    In contrast, shear instabilities and/or shocks can match the
    observed wind speeds of both planets. }
\label{fig:HD189}
\end{figure}

In this section we use the heat engine framework to predict how strong
winds would have to be to balance the two main proposed drag
mechanisms on hot Jupiters, namely magnetic drag and shear
instabilities. We then evaluate our predictions by
comparing them to observed wind speeds obtained from Doppler
spectroscopy.

For magnetic drag we combine \Eq{eq:Uray}
with a kinematic scaling for the effective Lorentz drag timescale
\citep{Perna_2010_1,Menou:2012fu,Rauscher_2013}. To be
  consistent with Section 3, we use $k_0=0.3$ in \Eq{eq:Uray}. The drag timescale is
\begin{equation}
\tau_\mathrm{mag} = \frac{4\pi H_e \rho}{B^2} \mathrm{,}
\label{eq:tauMag}
\end{equation}
where $B$ is the dipole field strength, $H_e$ the atmospheric
electrical resistivity, and $\rho$ the gas density. The electrical
resistivity is inversely related to the ionization fraction $x_e$,
$H_e \propto \sqrt{T}/x_e$, where $x_e$ is calculated from the Saha
equation \citep{Perna_2010_1}. For hot Jupiters the ionized gas is
largely potassium, for which we assume a solar abundance \footnote{For
  a planet with the equilibrium temperature of HD 209458b,
  $T_{eq} = 1450 \ \mathrm{K}$, the ionization fraction is
  $x_e=4.4 \times 10^{-11}$, which is much smaller than the solar
  abundance of neutral potassium and thus consistent with the
  approximations made in \citet{Perna_2010_1}. The corresponding
  magnetic resistivity is
  $H_e=2.0 \times 10^{14} \ \mathrm{cm}^2 \ \mathrm{s}^{-1}$.}.  We
expect that most dissipation occurs somewhere between the upper levels
probed by Doppler observations ($\sim10^{-3}$ bar) and the
photosphere, so we calculate winds over the range
$10^{-3} \leq p \leq 1 \mathrm{bar}$.  Note that Equation
\ref{eq:tauMag} does not include induced atmospheric fields. In
strongly ionized atmospheres induced fields can be significant
\citep{Rogers:2020,Rogers:2014,Rogers:2017a}, which means winds could
decrease faster with equilibrium temperature than implied by Equation
\ref{eq:tauMag}.

For shear instabilities we predict wind speeds analogous to
\Eq{eq:Unum}. We assume instabilities have a spatial extent $L$ and damp
the flow over a timescale $L/U$, so wind speeds scale as 
\begin{equation}
\label{eq:shearU}
U_\mathrm{shear} = k_1 \left(L \eta \sigma T_{eq}^4 \frac{g}{p}\right)^{1/3}.
\end{equation}
For consistency we use $k_1=1.1$, as in Section 3.
We note that Doppler observations probe the upper atmosphere only whereas our theory constrains large-scale
dissipation and thus should be representative of the bulk flow. Observable wind speeds could
potentially deviate from the bulk flow in atmospheres with large vertical shear. Nevertheless, we
expect that the comparison between our theory and observations is warranted,
given that a wide range of hot Jupiter GCMs produce equatorial jets
that are strongly vertically coherent
\citep{Showmanetal_2009,Heng:2011a,Liu:2013,Mayne:2014,Polichtchouk:2014,Cho:2015}.

Figure \ref{fig:HD189} compares the observed wind speeds
of $1.9^{+ 0.7}_{-0.6} \ \mathrm{km}~\mathrm{s}^{-1}$ for HD 189733b
\citep{Louden:2015} and $2 \pm 1 \ \mathrm{km}~\mathrm{s}^{-1}$ for HD
209458b\footnote{Note that these are $1\sigma$ error bars and the
  detection itself was only significant at $2\sigma$.} \citep{Snellen:2010} with our
theoretical predictions for the two drag mechanisms\footnote{We assume
  $p=1$ bar, $\eta = 0.2$, and $g=23\mathrm{m}~\mathrm{s}^{-1}$, with the last two
  values motivated by the phase curve amplitude and mass-radius
  measurements of HD 189733b.}. 
To indicate that our scalings aren't exact, the colored
  envelopes in Figure \ref{fig:HD189} reflect the dominant sources of
  uncertainty in our scalings. For magnetic drag the uncertainty is
  dominated by the pressure at which dissipation is assumed to occur,
  for shear instabilities we use the remaining mismatch between theory and GCM simulations\footnote{We conservatively use $100\%$ uncertainty (a factor of two) for winds
    predicted with \Eq{eq:shearU}.} in Section 3. Because the magnetic drag timescale is relatively
  sensitive to both temperature and pressure we additionally explored
  the impact of different pressure-temperature profiles, and find that
  most features in Figure \ref{fig:HD189} are robust (see Appendix).

  First, we find that the observations for HD 189733b can only be
  matched with a very strong dipole field of $\sim 100 \mathrm{G}$
  (Fig.~\ref{fig:HD189}, top panel).  Second, matching the
  observations for HD 209458b also requires a strong dipole field, on
  the order of $\gtrsim 30 \mathrm{G}$. Such a dipole is broadly in
  agreement with predictions from dynamo scaling laws for HD 209458b
  \citep{Yadav:2017}, which predict a dipole component at the poles of
  $\sim 50 \mathrm{G}$ (R. Yadav, personal communication).
  We conclude that magnetic drag is a plausible drag mechanism for HD
  209458b. In addition, given the potentially large uncertainties in
  both the Lorentz drag timescale (Equation \ref{eq:tauMag}) and
  dynamo scaling laws, magnetic drag cannot be ruled out for HD
  189733b, even though the required field strengths would be larger
  than currently expected. Further theoretical work could help reduce
  these uncertainties.  Our result that Lorentz forces are potentially
  unimportant for HD 189733b but may be important for HD 209458b
  therefore agrees with previous estimates that magnetic drag could
  become significant at $T_\mathrm{eq} \gtrsim 1400 \mathrm{K}$
  \citep{Menou:2012fu,Rogers:2014}.

In contrast to magnetic drag, we find that shear instabilities are a
plausible mechanism to match the observations of both planets
(Fig.~\ref{fig:HD189}, bottom panel). Our scaling predicts that wind
speeds increase moderately with $T_{eq}$, in agreement with the
observations. We also find that the vertical scale height $H$, which
has been proposed as the characteristic scale of Kelvin-Helmholtz
instabilities in hot Jupiters \citep{Goodman:2009,Li:2010}, would
yield wind speeds that are an order of magnitude too slow to match the
observed wind speeds. Instead, a damping length $2\pi a$, where $a$ is
the planet radius, is needed to match the observed wind speeds. Such a
damping length could be either due to a horizontal Kelvin-Helmholtz
instability or due to the steepening of day-night standing waves into
shocks. We note that the shock-resolving simulations in
  \cite{Fromang:2016} also found a dominant scale for horizontal shear
  instabilities of $L\sim 2 \pi a / 5$, and are thus consistent with
  our results. The upper end of our wind speed
  estimate is additionally consistent with the bulk flow becoming
  supersonic, and thus prone to dissipation via shocks (Fig.~\ref{fig:HD189}).

\section{Conclusion}
\label{sec:conc}

We describe the large-scale atmospheric dynamics of hot Jupiters by
modeling them as planetary heat engines. Hot Jupiters are comparable to, but less efficient than, ideal Carnot
engines because parcels lose heat
to space as they move between dayside and nightside.
Our theory successfully captures the intensity of winds in a large
number of hot Jupiter simulations (\Fig{fig:winds}). Remaining
differences between theory and simulations are likely due to our
imperfect understanding of numerical dissipation in the simulations,
instead of a fundamental shortcoming in our theory.

Applying our theory to observations, we find that either the magnetic
dipole field of HD 189733b could be stronger than current estimates
suggest, or that its atmosphere is dissipating kinetic energy via
shear instabilities and/or shocks.  For HD 209458b our results
indicate that both drag mechanisms can plausibly match the
observations.

Looking towards future observations, we expect that magnetic drag
should become dominant on hotter exoplanets with
$T_{eq} > 1400 \ \mathrm{K}$ (Fig.~\ref{fig:HD189}).  Wind speeds on
these planets should follow a different trend with equilibrium
temperature than wind speeds in colder atmospheres. As a result, we
propose that more Doppler measurements over a wider range of planets
could reveal a diversity of drag mechanisms at work in hot Jupiter
atmospheres.

\acknowledgements We thank Vivien Parmentier, Dorian Abbot, and Malte
Jansen for insightful feedback on an early draft. We also
  thank the reviewer for helpful comments that significantly improved
  this manuscript. This work benefited from the Exoplanet Summer
Program in the Other Worlds Laboratory (OWL) at the University of
California, Santa Cruz, a program funded by the Heising-Simons
Foundation. D.D.B.~Koll was supported by a James McDonnell Foundation
postdoctoral fellowship. T.D.~Komacek was supported by a NASA Earth
and Space Science fellowship.

\appendix

\section*{Sensitivity to numerical parameters}
\label{app:one}
 Our scalings suggest that, for simulations that are dominated by
numerical drag, large-scale wind speeds should be sensitive to
horizontal resolution (Eqn.~\ref{eq:Unum}).  To explore this
possibility we performed additional simulations in which we did not
include any Rayleigh drag (including no basal drag) and kept the
equilibrium temperature fixed to $1500 \ \mathrm{K}$ while varying
different numerical parameters in the model. The two parameters we
considered are the model's horizontal resolution and its timestep
$dt$. Table \ref{table:params} summarizes the numerical parameter
variations for this suite of simulations. The Shapiro filter timescale
$\tau_\mathrm{num}$ was always kept equal to the timestep.

\begin{figure}[h!]
\begin{center}
\includegraphics[width=0.5\textwidth]{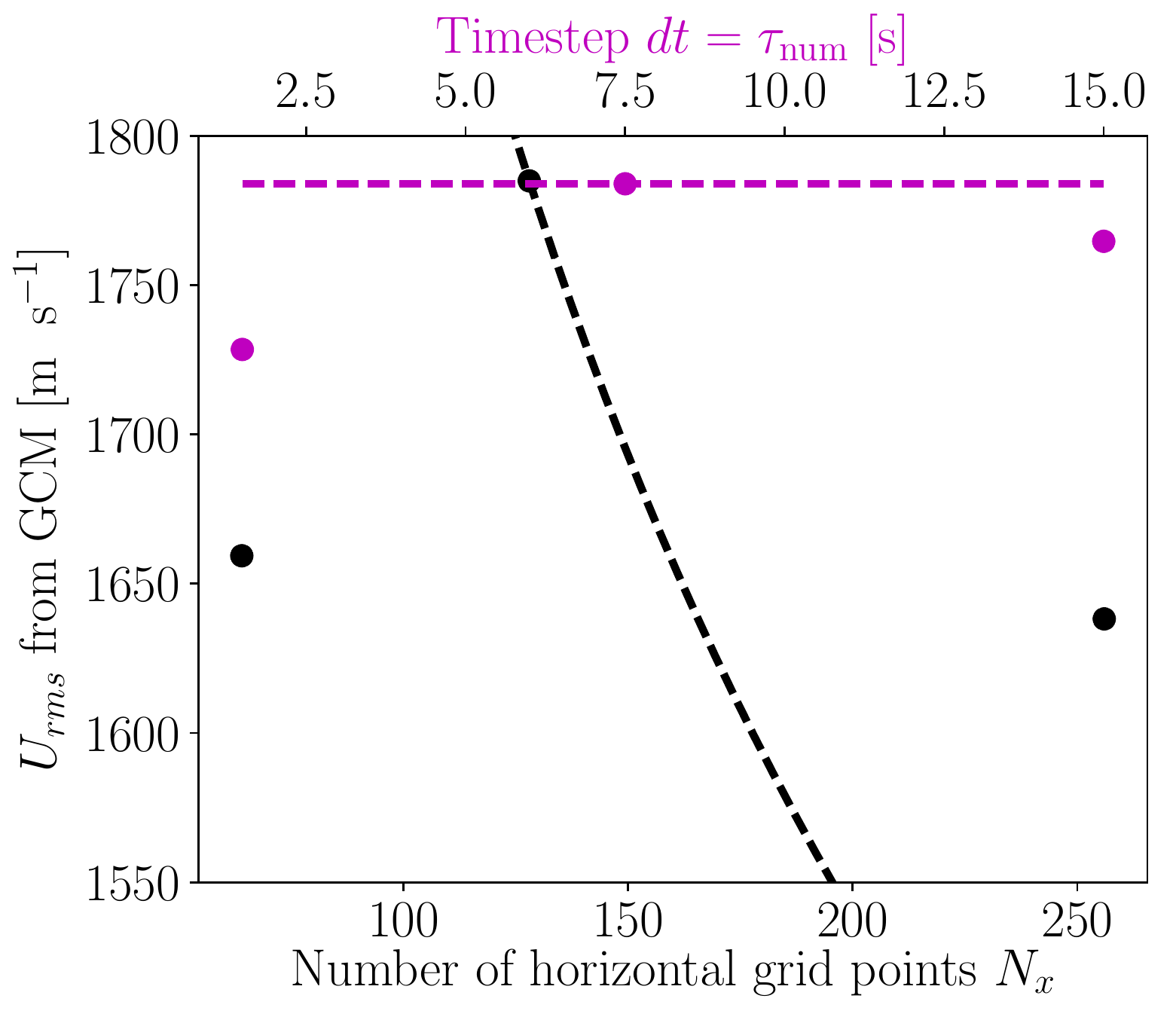}
\caption{Our scaling for how wind speeds depends on numerical
  parameters (Eqn.~\ref{eq:Unum}) matches the independence of
  $U_\mathrm{rms}$ on timestep well, but does not match the dependence
  of $U_\mathrm{rms}$ on grid size. Shown are GCM results for
  $U_\mathrm{rms}$ as a function of horizontal resolution (black dots)
  and timestep (magenta dots) from simulations with
  $T_\mathrm{eq} = 1500 \ \mathrm{K}$ and no Rayleigh drag. In this
  set of simulations the Shapiro filter timescale
  $\tau_\mathrm{num}$ is kept equal to the timestep. Dashed lines show
  our predicted dependence of $U_\mathrm{rms}$ on timestep (magenta)
  and resolution (black), using a value of $k_1$ such that the theory
  matches the intermediate GCM point. Eqn.~\ref{eq:Unum} correctly
  predicts that the wind speed is independent of timestep (accurate to
  the $3\%$ level in our GCMs), but predicts that the wind speeds
  should decrease steeply with increasing resolution, which is not
  found in our GCM simulations.}
\label{fig:numerics_scaling}
\end{center}
\end{figure}

\Fig{fig:numerics_scaling} shows that wind speeds are largely
independent of the GCM timestep. We only find a $\lesssim 3\%$
variation in the RMS wind speed while changing $dt$ (and thus also
$\tau_\mathrm{num}$) over an order of magnitude. Given that
Equation \ref{eq:Unum} predicts wind speeds should be independent of
$dt$, this implies a general agreement between our theory and our GCM
results.

In addition, Figure \ref{fig:numerics_scaling} shows that large-scale
wind speeds are less sensitive to horizontal resolution than our
scaling would suggest.  Following Equation \ref{eq:Unum}, wind speeds
should scale with resolution as $U \propto N_x^{-1/3}$, where $N_x$ is
the number of horizontal grid points. Our GCMs do not follow such a
scaling and instead we find that the wind speed is independent of
resolution to $\lesssim 10\%$ over a factor of 4 change in horizontal
resolution, going from C16 to C64.  One potential explanation is that
our weakly damped simulations develop a direct turbulent cascade of
energy to smaller scales, so that the large-scale kinetic energy of
the flow becomes insensitive to the dissipation scale. Another
explanation is that hot Jupiter GCM simulations are prone to
developing shocks
\citep[see][]{Rauscher:2010,perna_2012,Dobbs-Dixon:2013,Fromang:2016},
in which case the large-scale kinetic energy might be less sensitive
to how well the shock is being resolved than \Eq{eq:Unum} suggests.

Our result is consistent with the suggestion of \citet{Heng:2011a}
that changes in numerics can change wind speeds in GCMs at the
$\lesssim 10\%$ level, but shows that our scaling does not adequately
capture the dependence of large-scale GCM wind speeds on numerical
resolution.
As a result, a better description of numerical drag than our scaling
is needed to capture how hot Jupiter GCMs converge with numerical
drag. Nevertheless, although our scaling over-predicts
  the sensitivity to numerical parameters, it does correctly predict
  the sensitivity to physical parameters, such as equilibrium
  temperature (see \Fig{fig:winds}, left panel).

\begin{table}[h!]
\begin{center}
\begin{tabular}{| c | c | c |}
\hline
{\bf Physical Parameter} & {\bf Parameter Value(s)} & {\bf Unit} \\
\hline
Equilibrium temperature $T_\mathrm{eq}$ & 500, 1000, {\bf 1500}, 2000, 2500, 3000 & K \\
\hline
Visible absorption coefficient $\kappa_{SW}$ & $4 \times 10^{-4}$ & m$^{-2}$ kg$^{-1}$ \\
\hline
Thermal absorption coefficient $\kappa_{LW}$ & $2.28 \times 10^{-6}$ $\times \ (p/\mathrm{1~Pa})^{0.53}$ & m$^{-2}$ kg$^{-1}$ \\
\hline
Drag timescale $\tdrag$ & $10^3, 10^4, 10^5, 10^6, 10^7, {\bf \infty} $ & s \\
\hline
Gravity $g$ & 9.36 & $\mathrm{m} \ \mathrm{s}^{-2}$ \\
\hline
Rotation rate $\Omega$ & $2.078 \times 10^{-5}$ & $\mathrm{s}^{-1}$ \\ 
\hline 
Planet Radius $a$ & $9.43 \times 10^7$ & m \\ 
\hline 
Heat capacity $C_p$ & $1.3 \times 10^4$ & $\mathrm{J} \ \mathrm{kg}^{-1} \ \mathrm{K}^{-1}$ \\
\hline
Specific gas constant $R$ & 3700 & $\mathrm{J} \ \mathrm{kg}^{-1} \ \mathrm{K}^{-1}$ \\
\hline
\hline
{\bf Numerical Parameter} & {\bf Parameter Value(s)} & {\bf Unit} \\
\hline
Horizontal resolution ($N_x$) & C16 (64), {\bf C32 (128)}, C64 (256) & n/a \\
\hline
Vertical resolution $N_z$ & 40 & n/a \\
\hline 
Timestep $dt$ & 1.5, 7.5, {\bf 15} & s \\
\hline
Shapiro filter timescale $dt_\mathrm{num}$ & 1.5, 7.5, 15, {\bf 25} & s \\
\hline
Shapiro filter length scale $l_\mathrm{num} = 2\pi a/N_x$ & $2\pi a/64$, {\bf ${\bf 2\pi a}$/128}, $2\pi a/256$ & m \\
\hline
Shapiro filter order $n$ & 4 & n/a \\
\hline
\end{tabular}
\caption{Range of physical and numerical parameters used in our suite of simulations. Numerical parameters in bold show fiducial values used for our main suite of simulations with varying physical parameters, and physical parameters in bold highlight fiducial values used for our secondary suite of simulations with varying numerical parameters. Numbers in parentheses for horizontal resolution show the approximate number of horizontal grid points.}
\label{table:params}
\end{center}
\end{table}

\section*{Sensitivity of magnetic drag timescale to temperature-pressure profile}
\label{app:two}
 
\begin{figure}[h!]
\centering
\includegraphics[width=.5\textwidth]{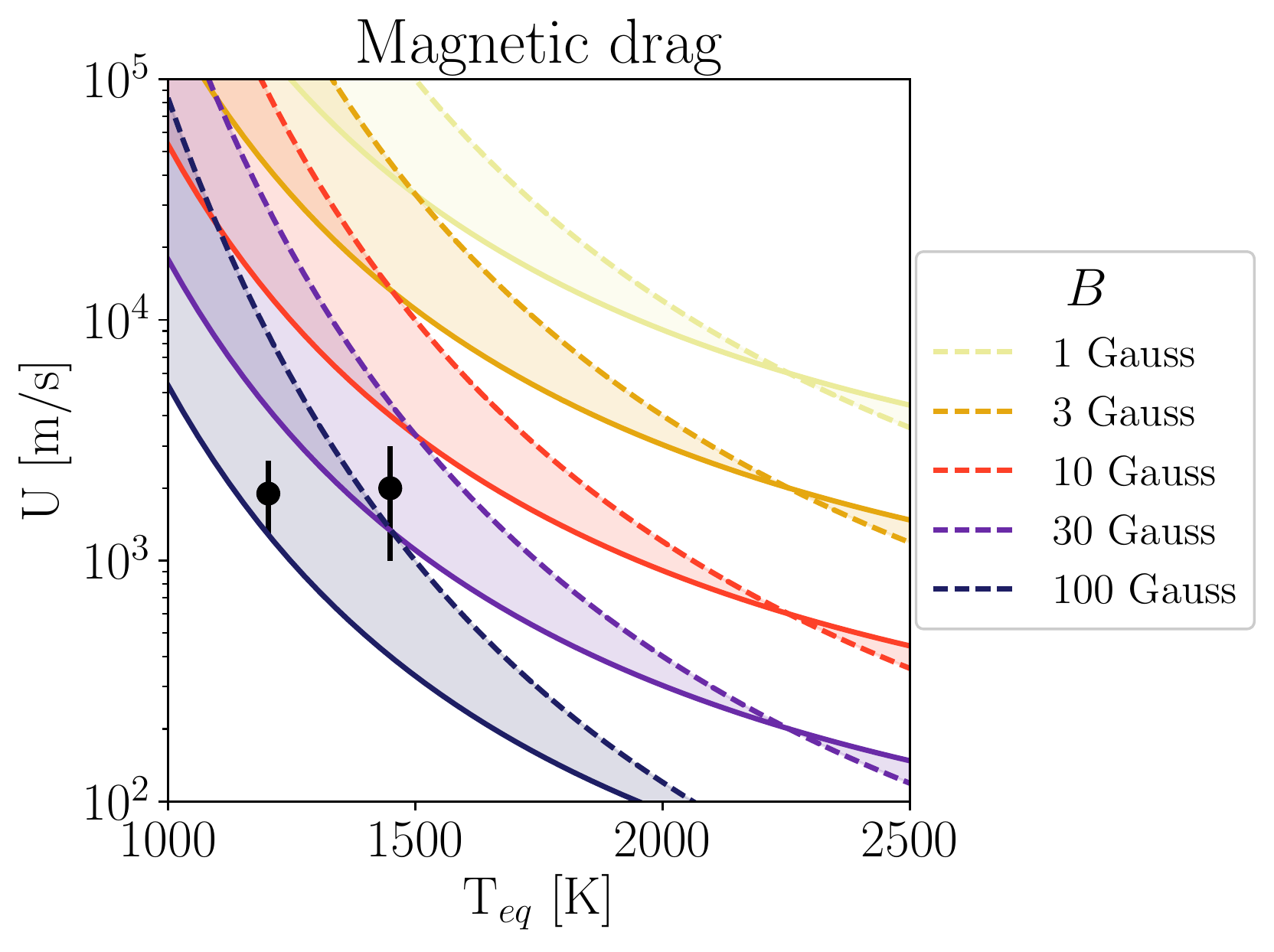}
\caption{Same as the top panel in Figure \ref{fig:HD189}, but instead
  of an isothermal atmosphere we assume that temperature increases
  with pressure following the analytic solutions in
  \citet{Guillot:2010}. Solid lines are evaluated at $1 \mathrm{bar}$,
  dashed lines are evaluated at $10^{-3} \mathrm{bar}$. Compared with
  Figure \ref{fig:HD189}, our main conclusions are robust to changes
  in thermal structure.}
\label{fig:HD189_appendix}
\end{figure}

Because the magnetic drag timescale is highly sensitive to
temperature \citep{Perna_2010_1,Menou:2012fu,Rauscher_2013}, we
explored the impact of the assumed temperature-pressure
profile on our results in Section \ref{sec:data}. In Section
\ref{sec:data} we assume an isothermal atmosphere, here we constrain
the vertical temperature structure using the analytical solutions from
\citet{Guillot:2010} as follows: we use Eqn.~29 from
\citet{Guillot:2010} with parameters similar to those used in that paper ($\kappa_{LW}=10^{-2}
cm^2 g^{-1}$, $\gamma=0.1$, $T_{int}=100\mathrm{K}$, $f=0.25$). With
these temperature-pressure profiles we evaluate the magnetic drag
timescale (Eqn.~\ref{eq:tauMag}) at $1 \mathrm{bar}$ and $10^{-3}
\mathrm{bar}$, and compute wind speeds following Eqn.~\ref{eq:Uray}.\\
\indent Figure \ref{fig:HD189_appendix} shows that our conclusions from
Section \ref{sec:data} are robust. The most significant difference in
Figure \ref{fig:HD189_appendix} compared to Figure \ref{fig:HD189}
occurs above $T_{eq}\gtrsim 1500 \mathrm{K}$, in which wind speeds
increase more slowly with temperature, whereas our scalings at
$T_{eq}<1500 \mathrm{K}$ are affected relatively little.  The
relatively small effect of the temperature-pressure profile is largely
due to a trade-off between the effect of pressure and temperature on
the magnetic timescale (Eqn.~\ref{eq:tauMag}). Although $H_e$ has an
exponential sensitivity to temperature, the absolute value of
temperature varies less than a factor of two between $1 \mathrm{bar}$
and $10^{-3} \mathrm{bar}$. This compares to a three order of
magnitude change in pressure, which appears in both density
($\rho \propto p$) and resistivity
($H_e \propto x_e^{-1} \propto p^{1/2}$) in Eqn.~\ref{eq:tauMag}.

\if\bibinc y

\fi

\end{document}